\documentclass[a4paper,onecolumn,9pt]{article}
\usepackage{eurosis}
\usepackage{url}
\usepackage{natbib}

\usepackage{graphicx}
\usepackage{gensymb}

\usepackage[left=1.62cm,right=1.62cm,top=1.9cm]{geometry}

\usepackage{amsmath}

\bibpunct[; ]{(}{)}{,}{a}{}{;}

\begin{document}

% vmi szimulacios specifikus dolgot meg beleirni par mondatot, pl h egyik valos ideju a amsik nem, a nem valos ideju meg akkora terkeppel meg forgalpommal nem tud boldogulni, amaz meg nem tud robotkarokat modellezni
% van industry 4.0 track is
\title{NETWORK RESOURCE MANAGEMENT FOR CYBER-PHYSICAL PRODUCTION SYSTEMS BASED ON QUALITY OF EXPERIENCE}

%hogyan teszek bele tobb author-t?
\author{Attila Vidács, Zalán Trombitás \\
HSN Lab, Department of Telecommunications and Media Informatics,\\ Faculty of Electrical Engineering and Informatics,\\ Budapest University of Technology and Economics, Hungary \\
email: \texttt{vidacs.attila@vik.bme.hu}
\and Géza Szabó \\
Ericsson Research, Budapest, Hungary\\
email: \texttt{geza.szabo@ericsson.com}
}

\date{}

\maketitle

\thispagestyle{empty}


\begin{abstract}
In today's industrial challenges, it can be observed that the trends point in the direction of agile, wireless connected robots where elements of intelligence and control are implemented in the edge cloud. %As a consequence of this, it has become an important question how network performance affects the quality of work performed by robots. 
This paper outlines the roles of three key participants in the value chain of an industrial process: the network provider, the robot operator, and the customer. It proposes a scheme where the Quality of Service (QoS) parameters of the robot are fed into the network to inform network resource management. A sanding process use case is simulated to demonstrate the relationship between QoS and Quality of Experience (QoE) for each participant, quantitatively.
A demonstration video is available at \citep{sanding-demo-video}.  
\end{abstract}

\section{INTRODUCTION}\label{intro}
Several industries aim to satisfy customers, which depends on various factors. The value chain is the sequence of activities involved in producing and delivering a product or service to the customer including suppliers, manufacturers, distributors, retailers, and service providers, each of whom adds value to the final product or service. The value chain leads to the customer's opinion of the final product. The customer satisfaction is the result of various factors, including the quality of the product or service, the performance of the transport network, and the customer's own preferences and expectations.

%Transporting digital content requires an exact replica of the original product at the delivery location. Telecommunication communities have resolved this issue using protocols and coding techniques for decades.
Streaming media marked the first step towards consuming digital goods. 
The quality of content that consumers perceive is subjective and relies on the content producer, transport network, and the consumer itself. Since the media transport process typically involves a direct relationship between the content provider and the end customer, allowing customers to set network requirements is a viable approach.

In Cyber-Physical Production Systems (CPPS), transmission channel properties, particularly those arising from under-provisioned networks, become materialized by creating unusable products with significant defects. To fix these physical products is both difficult and costly. Consequently, CPPS networks tend to be over-provisioned.
CPPS value chains consist of multiple distribution channels. The network requirements are often set by participants in the manufacturing process who are several hops away from the customer. We argue that the network requirements correlate with customer requirements, but each player in the chain assigns different weights to them. Typically, the participant operating the robots determines the network requirement based on the robots' operational characteristics.

In this paper we tackle the following main issues. 
We intend to experiment with the effect of network characteristics on the whole value-chain of a CPPS.
We elaborate on which Key Performance Indicators (KPIs) -- especially user experienced quality -- are important for a certain participant, and which should be finally taken into account as requirements on the network.
We define and propose robot control-specific QoS KPIs that are expected to correlate to the final (experienced) quality of the product. The aim is to have universal KPIs for a CPPS by being as product-agnostic as possible.
We utilize a detailed rigid-body simulation to validate our proposed scheme. The Scan'n'Plan \citep{scannplan} Robot Operating System (ROS) package is extended with our custom sanding simulation plugin for Gazebo \citep{Gazebo}. The plugin records the position, orientation, and timestamp of the polishing disc during each simulation run, models the sanding statistically and focuses on the Cyber-Physical Systems (CPS) aspect and its relationship with network effects.
The results of the simulation is utilized to provide input for our proposal of a network resource management scheme.

%Erdekes, nincs szamozva, nem lehet hivatkozni
%The paper is organized as follows. In Section~\ref{sec:relatedwork} we collect the relevant related work. In Section~\ref{sec:framework} we introduce the framework and the value chain model used for further analysis. In Section~\ref{sec:exampleusecase} we analyze the concept in an example use case of robotic surface processing. Section~\ref{sec:simres} discusses simulation results to support our proposal. In Section~\ref{sec:qocstrat} we discuss the relation to Network Resource Management (NRM). Section~\ref{sec:conclusion} concludes the paper.

%---------------------------------------------
\section{RELATED WORK}\label{sec:relatedwork}
%---------------------------------------------

The first step to standardize the assessment of subjective video quality (no other factors of overall experience) is done in ITU-T P.910 \citep{itup910}. 
It is accepted as the benchmark for evaluating subjective video quality with an output called Mean Opinion Score (MOS).
The MOS value is thus a numerical measure of the quality judged by people, often done on a scale from 1 to 5.
Objective quality algorithms (also known as objective models) are designed to mimic the behavior and perception of humans. The goal is to produce the same scores as MOS values, called estimated-MOS (eMOS). 

%There are three types of objective models: 1) no-reference models, where input is taken only from the receiving end of the media distribution chain using the most limited set of input parameters base the objective quality estimation on encoding rates, video resolution, frame rates, codecs and stalling information, as these factors provide the minimum amount of information about the video playout that is required to estimate a quality score; while 2) full-reference models can also be adopted, where the video originally transmitted is compared with the one that is received. 
%Another variant is the 3) reduced-reference model, where the original video is not needed for reference, but certain information about it is made available to the model.  
Quality assessment for industrial processes are defined in various standards (e.g.,~\citep{iso12944}, \citep{iso8501}) and manufacturer specific guidelines (such as \citep{surfaceval}). It is still an active research topic on how to make quality assessment more automated. It is out of the scope of this paper to define a subjective quality assessment method, but we make it one step closer by introducing objective KPIs that are directly correlated with subjective quality.

%Surface quality inspection is one of several significant operations in a system of industrial production. The majority of basic inspection techniques are carried out by trained inspectors, which can be time-consuming and difficult. Automated computer visual inspection techniques are introduced to enhance performance for industrial production with the advent of computer vision and artificial intelligence systems in \citep{chebrolu2022automated}.

%Authors of \citep{9618655} propose an autonomous sanding robot. The CAD model of the target object is automatically constructed with the structured-light technology, and the sanding behavior on the target surface is self-regulated under the desired impedance model. The suggested impedance controller has the benefit of being model-free because it uses adaptive neural networks to online correct for unclear dynamics and unidentified disturbances. The stability of the closed-loop system is rigorously proved with Lyapunov methods, and experimental results on different objects are presented to validate the performance of the developed robot. They focus on the control efficiency of the robot and assume that accurate control results in good sanding performance.
%ugyanez rövidebben:
Authors of \citep{9618655} propose an autonomous sanding robot. 
%The CAD model of the target object is automatically constructed, and the sanding behavior on the target surface is self-regulated. 
They focus on the control efficiency of the robot and assume that accurate control results in good sanding performance.
%Authors of \citep{8972597} present a collaborative human-robot framework for delicate sanding of complex shape surfaces. 
%A typical industrial manipulator that has a force/torque sensor and a specially created compliant control algorithm is used to perform delicate sanding.  Together with the compliant control, they discuss trajectory planning and safety problem of such an approach. 
%The experience of the human workers is exploited through the intuitive framework and applied to plan the trajectories for the robot.
Authors of \citep{sandingefficiencystudy} explored the correlation between the sanding efficiency and the surface quality of wooden materials, providing a method for judging the end of belt life in the production. 
They provide certain KPIs like surface removal rate or surface roughness, which are good QoE KPIs but does not say a final verdict on the MOS.
%Authors of \citep{analysisofsandingparams} analyse parameters of sanding with an abrasive sanding machine to reduce energy consumption and to improve processing efficiency and quality. The influences of grit size, feed speed, sanding speed, and sanding thickness on the sanding force, normal force,  arithmetic mean deviation of profile, power consumption, and power efficiency were analyzed by the orthogonal method in this study. 
%ugyanez rovidebben:
Authors of \citep{analysisofsandingparams} analyse parameters of sanding with an abrasive sanding machine to improve processing efficiency and quality. The influences of grit size, feed speed, sanding speed and sanding thickness were analyzed. These KPIs are difficult to relate to a network engineer who has little knowledge in sanding in such details. 
A higher level MOS score is preferred that is easy to comprehend and make actions in the network accordingly.
%Authors of \citep{sandingsimulation} proposed to automate polishing on 5-axis machining center using a passive elastomeric carrier. One of the main advantages of automatic polishing is the repeatability of the machine movements in order to achieve restricted form deviations. However, the material removal rate~(MRR) during polishing depends on parameters such as contact pressure, relative velocity and tool wear. In order to calculate the effective MRR throughout the polishing tool path with relation to the contact area and the contact pressure between the tool and the part, they consequently created a model specifically for their method. In this paper the simulation output is a heatmap with a detailed sanding model. We found this a good starting point for defining our eMOS.
%ugyanez rovidebben:
Authors of \citep{sandingsimulation} proposed to automate polishing on 5-axis machining center using a passive elastomeric carrier. One of the main advantages of automatic polishing is the repeatability of the machine movements.
%However, the material removal rate~(MRR) during polishing depends on parameters such as contact pressure, relative velocity and tool wear. In order to calculate the effective MRR throughout the polishing tool path with relation to the contact area and the contact pressure between the tool and the part, they consequently created a model specifically for their method. 
In the paper the simulation output is a heatmap with a detailed sanding model. We found this a good starting point for defining our eMOS.

%In \citep{cheng2017modeling} authors simulate the Abrasive Flow Machining (AFM) process and compare it with the real world process. Their finding is that the simulation can be considered detailed enough and acceptable to be applicable to the blade profile prediction and accuracy control. 

\section{FRAMEWORK}\label{sec:framework}
\begin{figure}[ht] 
\centering 
\includegraphics[width=0.36\textwidth]{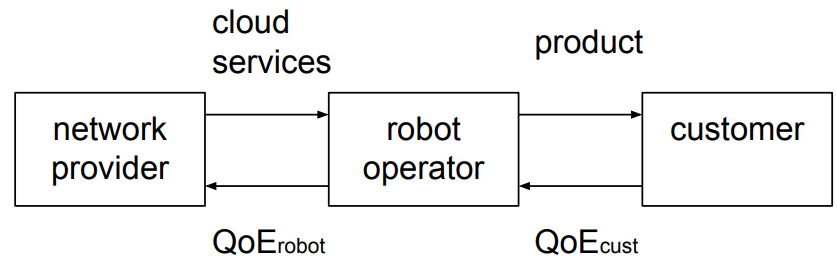} 
\caption{Players, provided services and experienced quality feedback.} 
\label{fig:gezalikesit} 
\end{figure}

In our study we consider the players constituting for the value chain of a simple CPPS as depicted in Fig.~\ref{fig:gezalikesit}.
%\subsection{Players}
The \textit{network operator} provides cloud and communication services to the \textit{robot operator}. The robot operator performs its cloud-based robot control that builds on that provided service. The final product that the robot produces are provided to the \textit{customer}. 

\subsection{Service provisioning and QoS}

The network operator manages its network and provides guaranteed cloud services where the service quality is described by a set of network-related QoS parameters ($QoS_{nw}$). The QoS must be set to satisfy the robot operator's needs and expectations.

The robot operator uses the cloud services to implement and operate its robot control. It's expectation is that the robot performs its tasks and be able to produce the final product with quality that satisfies its customer's needs. The product quality at the end of the line can be quantified by a set of QoS parameters that relates the final product only ($QoS_{prod}$).

The customer receives the final product from the robot operator.

\subsection{QoE and eMOS}
The customer decides whether the final product meets his/her expectations, and tries to express this overall level of satisfaction by a Quality of Experience ($QoE_{cust}$) value. This rating is, of course, in direct connection with product quality, but other (subjective) factors can also influence it. Hopefully, the overall customer satisfaction can be well estimated by the $eMOS_{cust}$ value that only depends on the (objective) product quality (see Eq.~\ref{eq:emos_cust} where $u(\cdot)$ denotes some utility function that takes into account and weights the different QoS parameter values). 
\begin{equation}
  eMOS_{cust} = u_{cust}(QoS_{prod})
  \label{eq:emos_cust}
\end{equation}
The robot operator can receive the $eMOS_{cust}$ value as a feedback, and use it to manage the production process, i.e., to set and guarantee the QoS for all elements of the production process, including the QoS required by the robot control method ($QoS_{robot}$). Some elements of the production process can have an impact on the final product quality, but have no connection with the cloud and networking services, such as resource quality of raw materials ($QoS_{materials}$) or used machinery ($QoS_{tools}$).
The first challenge is to find the correlation between the final product quality $QoS_{prod}$ and all the (possibly interdependent) QoS parameters of the production process, i.e., to find the function $g_{prod}$ in Eq.~\ref{eq:qos_prod}.
\begin{equation}
  QoS_{prod}= g_{prod}(QoS_{robot}, QoS_{materials}, QoS_{tools}, ...)
  \label{eq:qos_prod}
\end{equation}
Since the robot control is implemented using the cloud services provided by the network operator, the final product quality indirectly depends on the (experienced!) quality of those services.
The robot operator will be satisfied by the cloud services if and only if the robot control-related QoS parameters ($QoS_{robot}$) are adequate for achieving the desired product quality ($QoS_{prod}$) that will be satisfying for the customer. The satisfaction of the robot operator can be expressed by grading (e.g., using the $eMOS_{robot}$) the experienced cloud service quality according to Eq.~\ref{eq:emos_robot} where $u_{robot}$ is the utility function of the robot operator.
%555555555555
\begin{equation}
    eMOS_{robot}= u_{robot}(QoS_{robot})
    \label{eq:emos_robot}
\end{equation}
The network operator wants to provide cloud services to the robot operator that meets the robot operator's expectations. In order to do so, the network must be managed to provide various network QoS guarantees such as delay, jitter, loss, etc., denoted by $QoS_{nw}$. 
Assuming that the $eMOS_{robot}$ is provided for the network operator as a feedback, the challenge is to find out how the QoS metrics directly related to robot behaviour are connected with the cloud services, i.e., to find the function $g_{robot}$ in Eq.~\ref{eq:qos_robot}. 
\begin{equation}
  QoS_{robot}= g_{robot}(QoS_{nw})
  \label{eq:qos_robot}
\end{equation}
However, it is non-trivial whose task is to identify the function $g_{robot}$ in Eq.~\ref{eq:qos_robot}. There are different options for it, that will be elaborated in the next sections.

\subsection{Value chain}\label{sec:valuechain}

The robot operator is satisfied when its customer is satisfied with the end product. That is true for the network operator as well, it can only be satisfied if it's customer, the robot operator, was satisfied.
%$$QoE_{nw}\sim QoE_{robot}\sim QoE_{cust}$$
The challenge is to track down and correctly evaluate the situation when the end customer is not fully satisfied. The robot operator can not be fully satisfied either, but the dissatisfaction should not be blindly reflected when judging the received network service quality.
%
%ezt az abuse-ot igy hagyjuk? igen :) 
By combining Eqs.~\ref{eq:emos_cust}, \ref{eq:qos_prod} and \ref{eq:qos_robot} (with a slight abuse of notations) we have Eqs.~\ref{eq:emos_cust_dep} and \ref{eq:qos_prod_nw}.
\begin{equation}
\begin{split}
%eMOS_{cust} &= u_{cust}(QoS_{prod})\\
%&=u_{cust}(g_{prod}(QoS_{robot}), ...)\\
%&=u_{cust}(g_{prod}(g_{robot}(QoS_{nw}), ...))
eMOS_{cust} &= u_{cust}(g_{prod}(g_{robot}(QoS_{nw}), ...))
\label{eq:emos_cust_dep}
\end{split}
\end{equation}
Eq.~\ref{eq:emos_cust_dep} reveals that the customer satisfaction does depend on the network, but this dependence is rather hidden and complex.
\begin{equation}
\begin{split}
%QoS_{prod}&=g_{prod}(QoS_{robot}), ...)\\
%&=g_{prod}(g_{robot}(QoS_{nw}), ...)
QoS_{prod}&=g_{prod}(g_{robot}(QoS_{nw}), ...)
\label{eq:qos_prod_nw}
\end{split}
\end{equation}
Eq.~\ref{eq:qos_prod_nw} gives a more direct connection between product quality and network performance, but it is still challenging to rigorously formulate the interrelation.

\subsection{How to manage the network}

\begin{figure}[ht] 
\centering 
\includegraphics[width=0.48\textwidth]{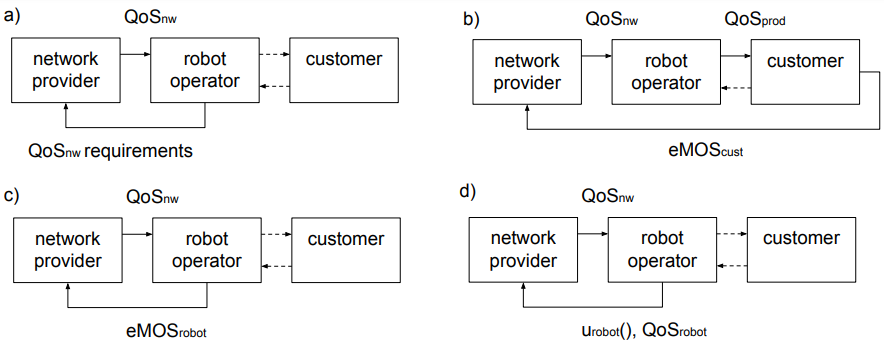} 
\caption{Options for providing feedback information for the network provider.} 
\label{fig:feedback-options} 
\end{figure}

The task for the network operator is to set up and manage the cloud services so that the provided service quality ($QoS_{nw}$) satisfies the needs. To be able to do that, information is needed on how the provided service is perceived by the robot operator. Different strategies can be applied based on what feedback information is available. 
Fig.~\ref{fig:feedback-options} lists four use cases that will be elaborated in the followings.

\subsubsection{Direct request from robot operator}\label{sec:directrequest}
This setup (see Fig.~\ref{fig:feedback-options}a) realizes the situation when the task of calculating the required network service quality guarantees is solved (i.e., $g_{robot}$ in Eq.~\ref{eq:qos_robot} is calculated) by the robot operator. To be able to do so, first it needs to know and understand the various network-related QoS parameters. The network measurements can be done either by the network operator or the robot operator, but these values must be available to the robot operator. After the calculations are made, the required QoS parameters are sent to the network operator who can configure its network to meet the needs.

\subsubsection{Direct feedback from customer}
In contrast to the direct request, the seemingly simplest and at the same time the most ambitious solution is when the network is managed to directly satisfy the customer (see Fig.~\ref{fig:feedback-options}b). When this goal is fulfilled the robot operator will be satisfied as well. However, the biggest problem with this approach is that the $eMOS_{cust}$ value depends on the network quality indirectly in more steps, see Eq.~\ref{eq:emos_cust_dep}. Another significant drawback is that the feedback is only available long after the production process ended and the product was handed over to the customer. We conclude that this scheme is unrealistic.
%vagy vmi further work kellene vagy vmi mashogy fogalmazni

\subsubsection{Simple feedback from robot operator}
The natural solution is to rely on the direct feedback from the robot operator when setting up the cloud services (Fig.~\ref{fig:feedback-options}c). The $QoS_{nw}$ values need to be tuned to maximize $eMOS_{robot}$ according to Eqs.~\ref{eq:emos_robot} and \ref{eq:qos_robot}. The connection is still indirect in two steps. Eq.~\ref{eq:qos_robot} gives how network quality influences the robot motion, and Eq.~\ref{eq:emos_robot} specifies what are the expectations of the robot operator regarding the actual production process. Considering a bit more complex manufacturing process, it is possible that the requirements for the robot precision changes from sub-task to sub-task even when the network quality ($QoS_{nw}$) and the corresponding robot quality ($QoS_{robot}$) do not change. 

The frequency of the $eMOS_{robot}$ feedback can vary. The value can be determined after finishing the product, or it can be calculated and sent periodically, or every at times when the production process enters a new phase. 

This feedback is simple, but its simplicity seems to be its main drawback as well. All details are compressed into a single value, making it difficult to extract and learn the possibly complex relationship between network quality and the needs and expectations of the robot operator.

\subsubsection{Detailed feedback from robot operator}\label{sec:feedbackfromrobot}
The idea here is to provide more detailed information for the network operator (see Fig.~\ref{fig:feedback-options}d).

When providing a detailed feedback on the measured robot-specific QoS parameters ($QoS_{robot}$), it makes it possible for the network operator to find the direct connection between network quality and robot behavior (see Eq.~\ref{eq:qos_robot}). Thus, a good estimate for the $g_{robot}$ in Eq.~\ref{eq:qos_robot} can be calculated. The next step is to find out the actual needs of the robot operator, especially when the requirements are changing because of the change in the production process. This can be communicated as well by directly providing the function $u_{robot}$ in Eq.~\ref{eq:emos_robot}. (Practically, this function is like a weighted sum of all robot-specific QoS values, where the weights represent the importance of each component regarding the actual sub-task carried out next.) 

When the feedback is provided frequently, or even in real-time, that would make it possible for the network operator to adaptively adjust the network QoS parameters according to the actual requirements.

\section{EXAMPLE USE CASE}\label{sec:exampleusecase}

The example use case on which we evaluate our feedback concept is a fully automated robotic surface machining where a robotic arm grinds, paints and polishes certain product.
%\subsection{Product quality}
%We assume that a set of QoS KPIs (denoted by $QoS_{prod}$) can be identified to quantify the product quality.

\subsection{Customer satisfaction}
The customer is given the final sanded, polished and painted product that he/she evaluates (see Eq.~\ref{eq:emos_cust}). His/her final judgement is primarily based on the product quality parameters ($QoS_{prod}$) but they are weighted according to personal preferences ($u_{cust}$). 
For example, the customer is not fully satisfied either because 1) the product has the wrong color shade, or 2) the product surface is spotted or scratchy.
%, or 3) she/he is not in the right mood on a Monday morning. 
%The customer's attitude can not be influenced by the management of the production process.
The color of the product depends on the choice and quality of the paint ($QoS_{material}$), but has nothing to do with robot control.  Only the product surface quality can be directly influenced by quality of robot control including the robot tools used. The first parameter set ($QoS_{robot}$) has a direct connection with the network service quality, while the second parameter set ($QoS_{tool}$) is independent of the cloud services used.

%We assume that the customer's utility function $u_{cust}$ values the product quality according to Eq.~\ref{eq:emos_cust}, where the product quality can be measured objectively as in Eq.~\ref{eq:qos_prod}.

\subsection{Robot operator's satisfaction}
To determine the $eMOS_{robot}$ value in Eq.~\ref{eq:emos_robot}, measurable QoS parameters are needed. 
As a first step, it should therefore be established which aspects are dominant in the case of each sub-process of the work process. This can be determined by comparing the quality metrics of the finished products and the monitored robot-specific QoS metrics.
It is also necessary to determine how important the metrics are. The ultimate goal would be to determine a formula ($u_{robot}$) that could be substituted into Eq.~\ref{eq:emos_robot} to get the final eMOS value.
%The utility function must be set so that the final eMOS value is at least 1 and at most 5.

In order to determine the dominant metrics, their weight and tolerance, real measurements are needed. The correlation between the individual measurements and the quality of the work piece could be found empirically, and the weights can be determined from the degree of correlation.

\subsection{Robot control-specific quality metrics}
Here we propose simple and measurable QoS metrics ($QoS_{robot}$) that can describe the robotic machining where a robotic tool is guided along a planned trajectory, like in surface treatment such as sanding, painting and polishing.

%\subsubsection{Trajectory error}
%\paragraph{Trajectory error.}
\textit{Trajectory error:}
The work performed can be considered successful if the planned trajectory was followed closely by the work tool. The trajectory error measures the deviation between the planned and actual trajectories that the robot tool traverses. 
%However, determining the degree of deviation from the planned route is non-trivial, since ...
%The trajectories consist of waypoints with associated time stamps stored sequentially that can be compared. 

\textit{Velocity:}
It is also important how the instantaneous velocity of the tool changes, how much acceleration and deceleration it experiences. So the metrics might be the mean, maximum, minimum, and the standard deviation of the tool speed.

\textit{Z distance:}
The perpendicular distance quantifies how much the robot arm keeps the work tool at a constant height from the machined surface. The metrics could be the largest and/or mean deviation from the desired height above the processed surface.

\textit{Orientation error:}
Another factor affecting the product quality may be the perpendicularity of the work tool to the surface. 
%The device may swing when changing direction, especially in the case of large velocity changes, so it is worth observing the completed tool trajectory from this point of view as well.
The roll, pitch and yaw values describing the orientation are to be measured.

\subsection{Network service quality}

The service quality ($QoS_{nw}$) provided by the network operator can be well measured by the classical QoS parameters such as end-to-end communication delay, jitter, bandwidth, error rate, etc. One of the most important network service quality parameter among all is the delay. 
%Note that de-jittering can be performed by hold and forward functionality resulting in added delay. Bandwidth is not significant for control messages.

% Gazebo -> \citep{1389727}; ROS->\citep{ros}  
\section{SIMULATION SETUP}\label{sec:simsetup}

\begin{figure}[t] 
\centering 
\includegraphics[width=0.38\textwidth]{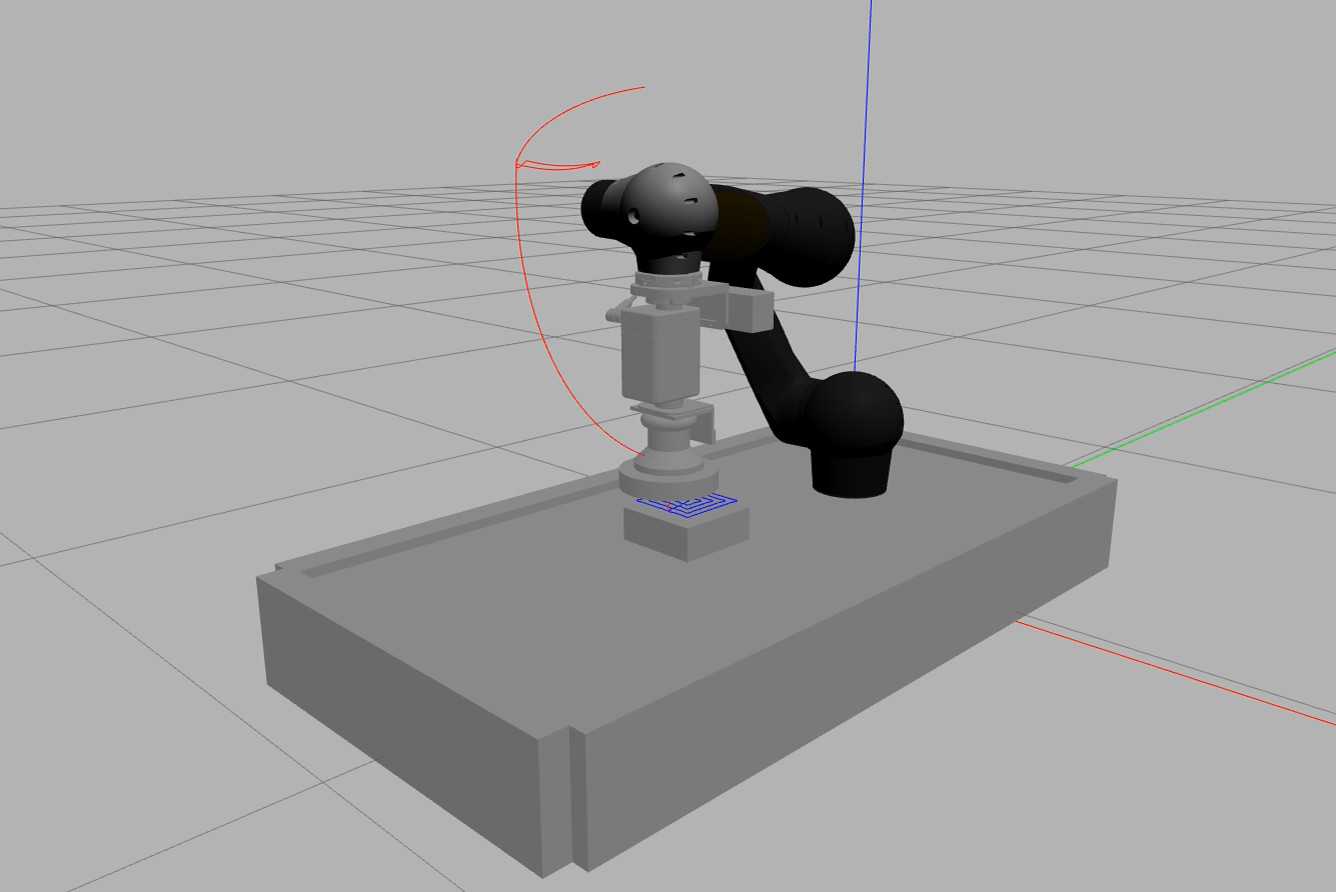} 
\caption{The simulation of robotic sanding in Gazebo} 
\label{fig:sanding}
\end{figure}

We conducted ROS/Gazebo simulation tests on a robot performing surface machining work.
%Gazebo \citep{Gazebo} is a 3D physical simulator for CPS systems, while ROS serves as an interface for robots and robotic system components. The combination of the two results in a powerful robot simulator. 
We used the Scan and Polish implementation \citep{scannplan} that includes a polishing robot arm with a laser scanner and a sanding tool, that is controlled remotely using velocity controls (see Fig.~\ref{fig:sanding}). %This implementation was created in the ROSIN project \citep{rosin}. 
Our delay plugin \citep{gazebopluginconf} can add a delay to the communication between ros\_control and Gazebo. The ros\_control package is a set of packages containing control interfaces, control handlers, transfers and hardware interfaces. 
%The plugin consists of two parts. The first is robot\_hw\_sim\_latency, which is a modified version of the default\_robot\_hw\_sim plugin of the base gazebo\_ros\_control, and this allows you to load custom plugins that can add latency to the communication between ros\_control and Gazebo. And the second one is latency\_plugin\_simple\_queue which is a simple latency plugin that uses queues to implement latency.
%In the simulation environment, in addition to the one implementing the delay, one more plugin was needed. This had two tasks, one is to save the coordinates to a file in case of suitable conditions, and the other is to visualize the path taken by the polishing wheel in the simulation
%...the simulation: scanning phase and polishing phase 

%The robot arm with a laser scanner and a sanding tool is part of the virtual production cell environment and is controlled remotely using velocity controls (see Fig.~\ref{fig:sanding}). 
%Traditional programming for robots calls for an expert-in-the-loop. This reduces the ability to adapt to process changes. By removing the human from the process and enabling agility, the Scan-N-Plan approach uses 3D scanning techniques to generate the part shape and position on-the-fly, in real-time. The trajectory planner uses the part geometry as a direct input to design unique trajectories specifically for the scanned objects. 
%Fig.~\ref{fig:sanding} shows the simulation screenshot of the final state of the sanding process. 
The sanding is performed fully automatically by the robotic arm, starting from surface scanning and trajectory planning in the first phase, and executing the actual sanding in the second phase. 

%The robot's laser scanner can be triggered to scan an object's surface.
%The scanned results are used to plan the sanding trajectory.
%in the following way:
%To depict the paths for sanding, it first generates a sequence of polygons; then, the polygons are arranged so that the path starts at the innermost point, spirals out to the outermost point, jumps to the next incomplete inner, and then spirals out to the greatest incomplete outer.
%The robot arm then follows this intended course. 

%ezt a plugin dolgot reszletesebbre vettem
We implemented a Gazebo plugin to have input for modelling the sanding. 
During each simulation run, the plugin saves the x, y and z coordinates, roll, pitch and yaw orientation values of the polishing disc of the robotic arm are saved, together with the time stamp during polishing.
The resolution of this data is in the magnitude of 0.5~mm between two stored points.
The trajectory of the tool is visualized and can be tracked throughout the whole sanding process in the simulation.
%We do not go into the field of accurate simulation of the surface during sanding, but model that part and focus on the CPS part and its relation to the network effects.
We modelled the sanding as removing particles from a surface according to a two-dimensional normally distributed random variable $\{X, Y\}$ with zero mean and standard deviation set by the tool-size resulting in different sanding imprint for the various sanding strategies.
The imprint of the sanding tool is moved along each saved trajectory coordinate. The number of hits on a certain coordinate of the imprint is cumulatively added to the specific coordinate forming a heatmap.

\section{SIMULATION RESULTS}\label{sec:simres}
The following section discusses the results of the simulation.

\begin{figure}[t] 
\centering 
\includegraphics[width=0.16\textwidth]{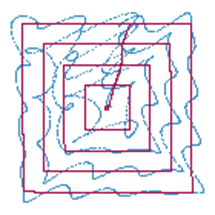} 
\includegraphics[width=0.16\textwidth]{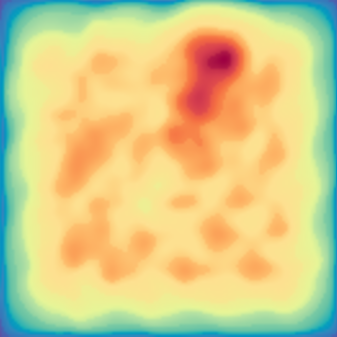} 
\caption{Planned (red) and real (blue) path for 66 ms delay (left), and heatmap of product quality (right).} 
\label{fig:trajectory-error}
\end{figure} 

\subsection{Product quality}
Our goal was to construct a simple measure to quantify the final product surface quality ($QoS_{prod}$). After modeling the sanding process in the simulation, the outcome of a simulation trial is a two-dimensional heatmap where the values represent the deviation of the surface level from the local average (see Fig.~\ref{fig:trajectory-error} right).
Next, the value known as the Earth Mover's Distance (EMD) is calculated for the heatmap.
The best example to illustrate the value of EMD is the following: we want to know how much work we need to do in the event that we want to move a pile of earth to another position, to a different arrangement, with the least amount of work possible. The algorithm only works if the amount of land is the same for both piles of land \citep{710701}.
Since the EMD sums up all the "bumps and pits" on the surface of the product, the higher the EMD value, the worse the quality of final product is.

\begin{figure}[t] 
\centering 
\includegraphics[width=0.49\textwidth]{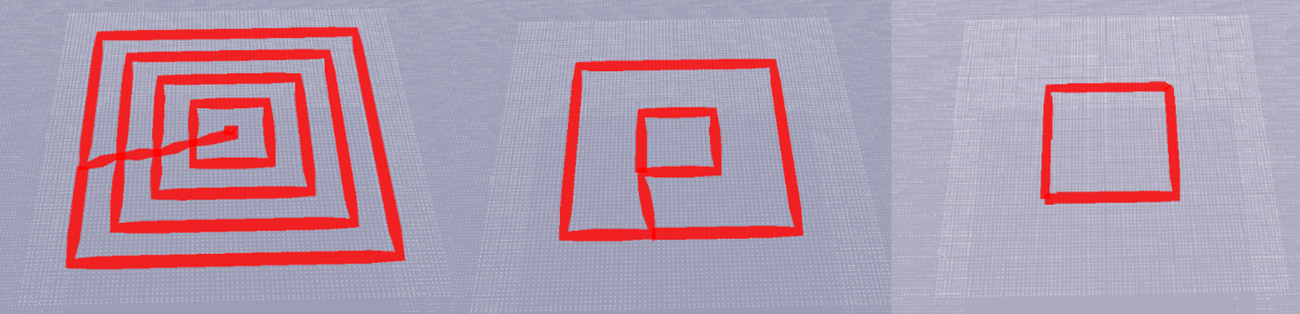} 
\caption{Planned trajectories for tool sizes of 12.5, 25 and 37.5 cm, respectively.} 
\label{fig:trajectories} 
\end{figure} 

\begin{figure}[t] 
\centering 
\includegraphics[width=0.49\textwidth]{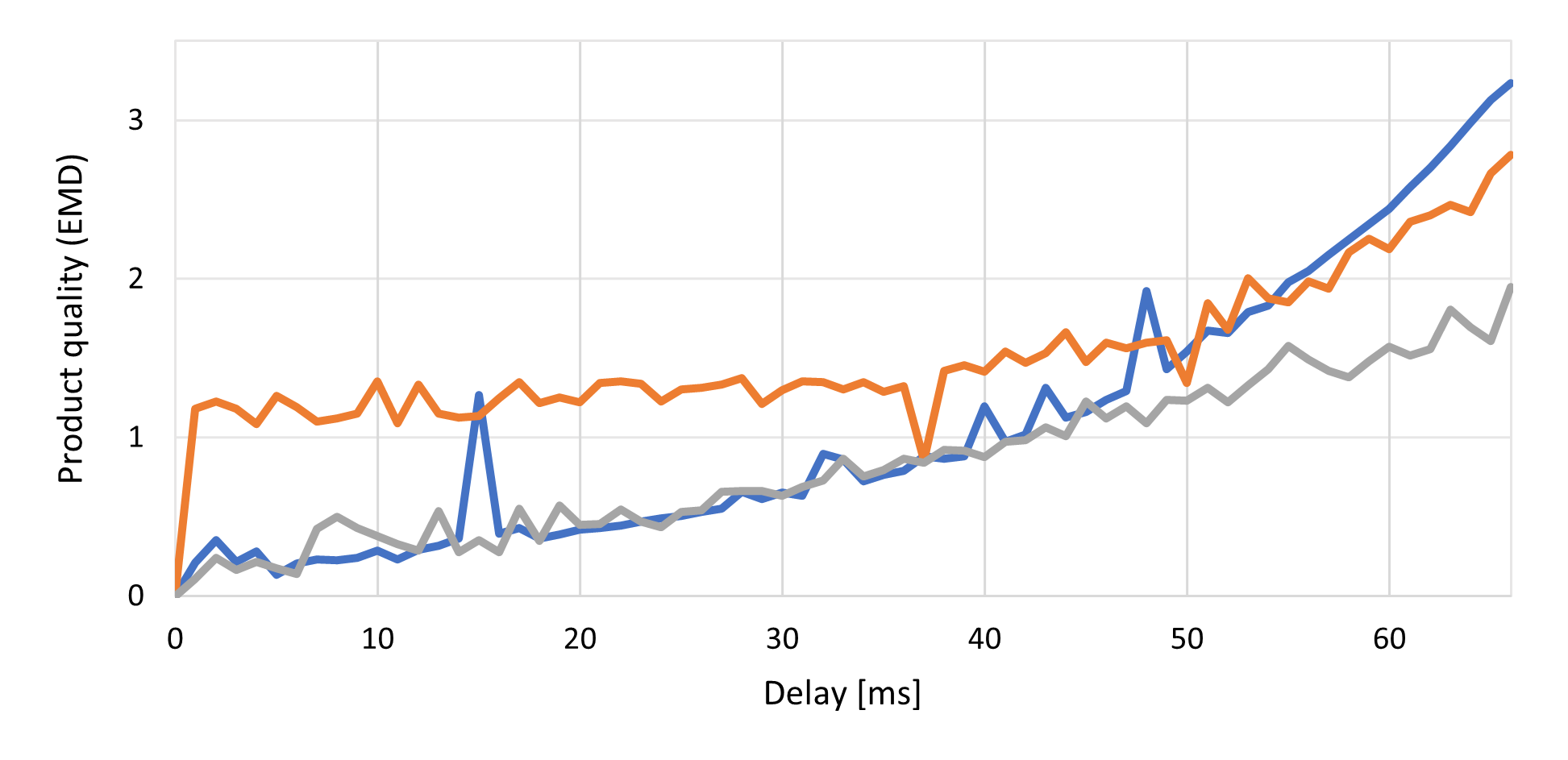} 
\caption{Product quality as a function of delay for different tool sizes: radius of 12.5 mm (blue), 25 mm (orange) and 37.5 mm (gray).} 
\label{fig:emd-delay}
\end{figure}

%\subsection{Tool size}
To illustrate that the final product quality can depend on many factors besides the network quality, simulations were performed for three different tool sizes: the radius of the polishing disc took values of 12.5, 25 and 37.5 millimeters. (The related planned trajectories are depicted in Fig.~\ref{fig:trajectories} for illustration.)
Fig.~\ref{fig:emd-delay} shows how the product quality changes for different tool sizes as the network delay increases. Interestingly, for small and medium tool sizes the final product quality is the same for various network delays up to a certain threshold (40~msec), but above the threshold the quality drops rapidly for the small tool size. Surprisingly, the medium size tool performs much worse for small delays, although it seem to be more robust when the network delay increases significantly.

As Eq.~\ref{eq:qos_prod_nw} shows, the final product quality eventually depends on the network performance, but there can be other factors -- unrelated to network quality -- that significantly affect product quality as well (see Eq.~\ref{eq:qos_prod}). So the challenge is to find those robot control-related QoS parameters ($QoS_{robot}$) that capture all the network dependent effects, and examine their relation to the final product quality.

%In the current investigation, we take into account the radius of the disc in the simulation. In the new measurements, the radius of the disc took values of 12.5, 25 and 37.5 millimeters.

%It is very interesting that for the smallest and largest radius, the EMDs take approximately the same values up to 47 milliseconds, and the slope of the increase is also roughly the same. After that, however, the values belonging to the smaller radius start to increase more steeply. And the diagram for the 25 millimeter radius is completely different from the other two. This already jumps to a value of around 1.2 at a delay of 1 millisecond, and then its value does not change much until 50 milliseconds, it jumps around 1.2. In this section, this function also takes the largest values. And in the section above this, its values also begin to increase, to a similar extent as the diagram for the 12.5 millimeter radius.

%The 12.5 millimeter disc radius was used in the followings. 

\subsection{Robot control quality}
\textit{Trajectory error:}
On Fig.~\ref{fig:trajectory-error} we can see the planned and real path for a delay of 66~msec. It can be seen that with such a delay, the polishing wheel can only follow the planned trajectory very poorly.
\begin{figure}[t] 
\centering 
\includegraphics[width=0.49\textwidth]{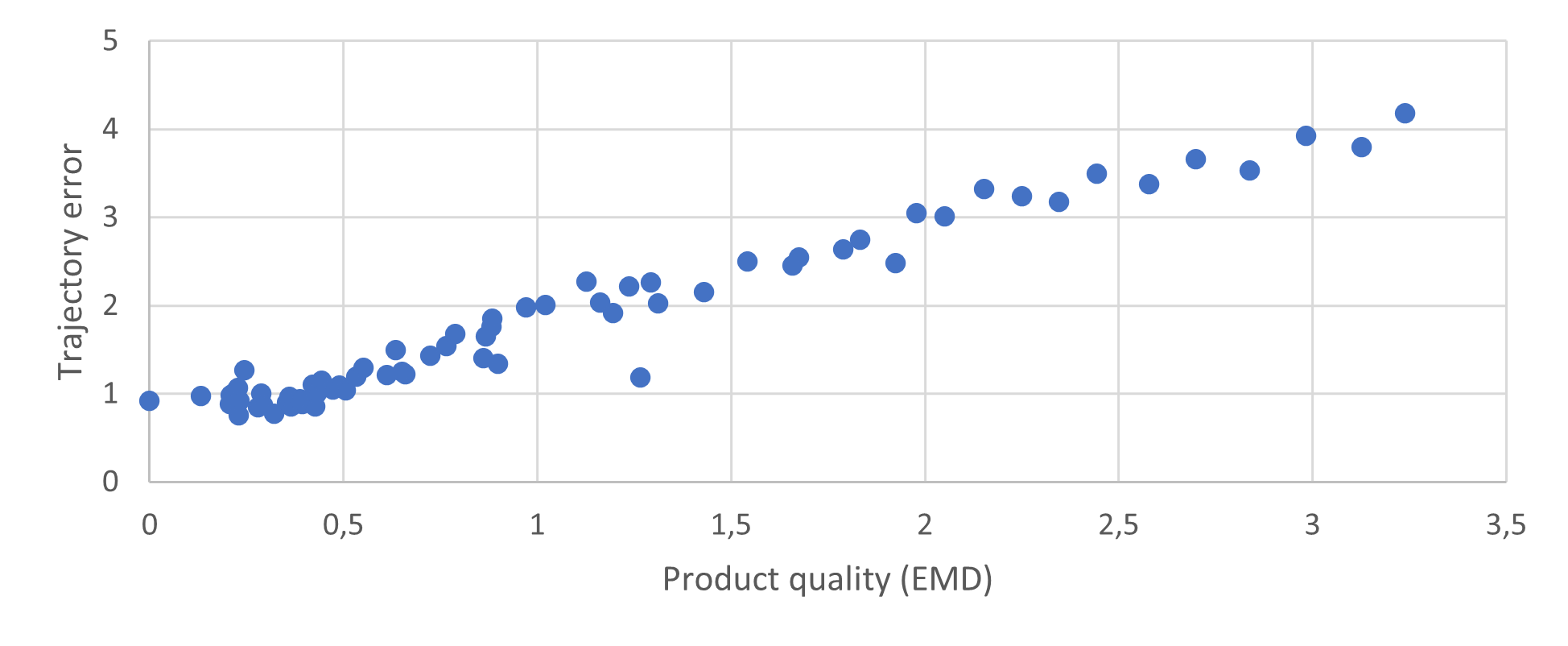} 
\caption{Deviation from the planned route depends on the quality.} 
\label{fig:trajectory-emd}
\end{figure}
It can be clearly seen on Fig.~\ref{fig:trajectory-emd} that as the product quality decreases the actual trajectory that the robot tool traverses deteriorates more and more from the planned one. Thus, this metric correlates with the change in quality, so it can be considered a dominant QoS metric.

\begin{figure}[ht] 
\centering 
\includegraphics[width=0.49\textwidth]{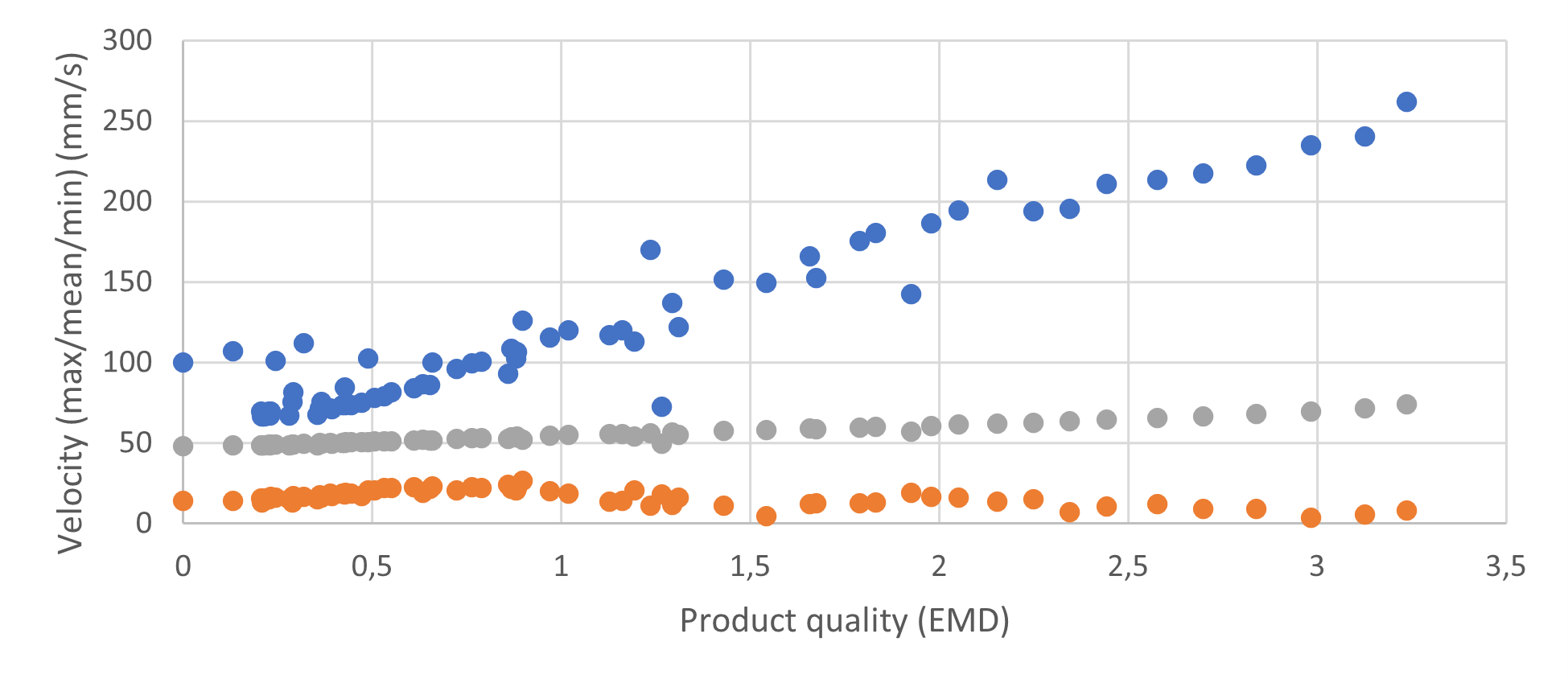} 
\caption{Tool maximum (blue), average (gray) and minimum (orange) velocity as a function of product quality.} 
\label{fig:velocity-emd}
\end{figure}

\textit{Velocity:}
Fig.~\ref{fig:velocity-emd} shows the change of tool velocity as a function of the quality.
The maximum and average velocities show an increasing trend with the decrease of work quality, so they are suitable measures for characterizing the robot performance quality.

\begin{figure}[t] 
\centering 
\includegraphics[width=0.49\textwidth]{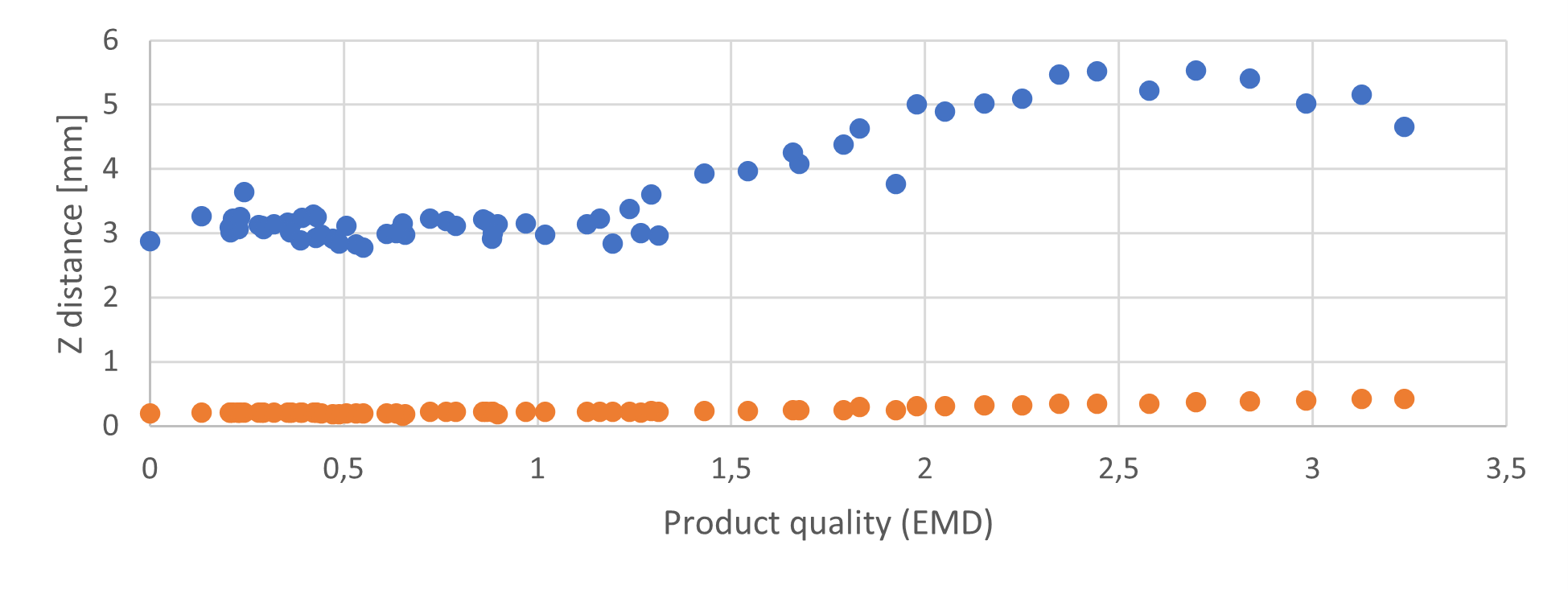} 
\caption{The max (blue) and mean (orange) Z distance as a function of quality.} 
\label{fig:z-distance-emd}
\end{figure}

\textit{Z distance:}
Fig.~\ref{fig:z-distance-emd} shows the Z distance as a function of product quality. For high quality products (i.e.,~EMD less than 1) the measured maximum Z distance must be kept around 3~mm in the current situation. As the value increases the quality deteriorates. However, there is no clear linear trend between the Z values and product quality.

\textit{Orientation error:}
We found that the change in perpendicularity is not proportional to the deterioration of the quality, thus the change in perpendicularity is not a dominant aspect for our case. (The figure is omitted here due to lack of space.) 

%It can be seen from Fig.~\ref{fig:perpend-emd} that the change in perpendicularity is not proportional to the deterioration of the quality. All three metrics move around the initial value, there is no clear downward or upward trend. 
%However, it is very interesting that there are "interruptions". The largest ones occur at the maximum deviation, but it can be seen that they occur in the same places for all three characteristics. The reasons for these are still unknown to me, but it is an interesting question, in any case, why they occurred.

%It should also be taken into account that the deviations are millimeter values, even though the two unit vectors, whose endpoints were the difference, are both one meter long. So we can see that these deviations from full perpendicularity are very minimal. 
%In summary, it can be said that in this simulation, the quality is not well characterized by the change in perpendicularity. Thus, it is not a dominant aspect and none of the possible metrics can be used.

%\begin{figure}[ht] 
%\centering 
%\includegraphics[width=0.44\textwidth]{figures/perpend-emd.png} 
%\caption{The maximum (blue) and mean (orange) perpendicular deviation as a %function of quality.} 
%\label{fig:perpend-emd}
%\end{figure}

\subsection{Experienced cloud service quality}
A robot operator can express its opinion on received network services by giving its utility function ($u_{robot}$) and listing the KPIs ($QoS_{robot}$) that are most valuable for the production process (see Eq.~\ref{eq:emos_robot}). 
In our simulation experiments the identified key quality parameters are the trajectory error and tool velocity (mean and max) with high weights in the utility function, optionally the Z distance, but not the orientation error.

The robotic arm is used for both the scanning and the sanding phases, so the network could influence both phases. Thus, the utility function should also indicate which processing phase is the actual phase. 

In our specific example the scanning is performed by first moving the laser scanner into four fixed positions above the work piece. When the laser scanner arrives and stops there, it creates point-clouds, collects them and performs the surface detection. As the scanning is performed when the robotic arm is already stopped, the network effect has no influence on the quality of surface modeling, assuming that the point cloud data can be transferred for cloud processing. This behavior can be indicated by the utility function by not giving any strict requirements on robot KPIs that are only relevant for the sanding phase. (Note, that if the application was scanning while moving, the network latency would certainly had an impact on the final scanned results.)

%\subsection{$QoS_{nw}$}
%...Fig.~\ref{fig:trajectory-delay}...

%Fig 11 az kb. alapvetes
%Fig 12-et meg nagyon ertem egyelore

%\begin{figure}[t] 
%\centering 
%\includegraphics[width=0.44\textwidth]{figures/trajectory-delay.png} 
%\caption{Deviation from the planned route depends on network delay.} 
%\label{fig:trajectory-delay}
%\end{figure}

%\begin{figure}[t] 
%\centering 
%\includegraphics[width=0.44\textwidth]{figures/velocity-delay.png} 
%\caption{Maximum (blue), average (gray) and minimum (orange) tool velocity as a function of network delay.} 
%\label{fig:velocity-delay}
%\end{figure}

\section{THE PROPOSED NETWORK RESOURCE MANAGEMENT SCHEME}\label{sec:qocstrat}
In order to allow factories (and other verticals) to automate the system integration and 5G network configuration tasks, 3GPP introduced the Service Enabler Architecture Layer for Verticals~(SEAL) in 3GPP Release~16.  
%The 3GPP Technical Specification Group~(TSG) Service and System Aspects~(SA) WG6~(SA6) is the application enablement and critical communication applications group for vertical markets. The goal of SA6 is to provide application layer architecture specifications for 3GPP verticals, including architecture requirements, functional architecture, procedures, information flows, interworking with non-3GPP application layer solutions, and deployment models as appropriate. 
3GPP TS~23.434 \citep{3gppTS23434} specifies APIs for various network functionalities including network resource management (NRM). % provisioning, connection management, device management, connection monitoring, group management, user profile retrieval, identity and key management, location reporting, events, and . 
%The above APIs are used to set up a certain production cell for a normal operation. If the production cell is reconfigured, then the existing communication resources are deleted and re-initiated. The operation does not require reconfiguration of the resources dynamically. 
This standard supports the "Direct request from robot operator" mode (see Section~\ref{sec:directrequest}).
In order to get to the "Detailed feedback from robot operator" mode (see Section~\ref{sec:feedbackfromrobot}) an extra node is required that does the translation from the utility function and the robot control quality parameters ($QoS_{robot}$) to network QoS, as given by Eq.~\ref{eq:g_nw} that is based on Eqs.~\ref{eq:emos_robot} and \ref{eq:qos_robot}.
\begin{equation}
  QoS_{nw} = g_{nw}(u_{robot}(), QoS_{robot})
  \label{eq:g_nw}
\end{equation}

Architecture-wise this can happen in either the network operator's or the robot operator's domain.
Assuming that such measurement that is presented in Section~\ref{sec:simres} is available at the robot operator, and the operation mode of the existing SEAL standard, the \S 14.3.2.13 defined "end-to-end QoS management request" from the NRM client to the NRM server should be filled with the following information elements: "list of VAL UEs" for whom the end-to-end QoS management occurs, the "IP address" of the VAL UE and the "end-to-end QoS requirements" of the application including latency, error rate, etc. for the end-to-end session.
The sanding process with a good enough quality requires trajectory error smaller than 3, tool max velocity less than 150~mm/s which requires product quality EMD to be less than 1.5 (see Fig.~\ref{fig:trajectory-emd}, \ref{fig:velocity-emd}) induces 40~ms latency requirement with a 25~mm tool size assuming no jitter. %As the jitter can be transformed to jitter, in case of a 
The bandwidth requirement of the robot control process is in a magnitude of 1~Mbps.

\section{CONCLUSIONS}\label{sec:conclusion}
%------------------------------------------
In this paper, we examined the key participants in the industrial process value chain, including the network provider, robot operator, and customer. We explored their relationships and various approaches to managing network resources. To analyze network resource management for a remotely controlled sanding process, we developed a simulation setup. The example provided quantitative demonstrations of how Quality of Service and Quality of Experience are linked for different participants. Using the simulation results, we proposed a schema in which the network receives QoS parameters and their utility function from robot control. With this information, the network operator can calculate and establish the necessary QoS parameters for the specific use case.

As a future work, there are numerous other industrial processes that can be thoroughly assessed to determine their performance in relation to the network QoS within a CPPS setup.

\bibliography{refs}

\end{document}